\documentclass[12pt,cites,graphics]{iopart}

\usepackage{graphicx}

\begin{document}

\title{Phase behaviour of charged colloidal sphere
dispersions with added polymer chains}
\author{Andrea Fortini$^{1}$, Marjolein Dijkstra$^{1}$ and Remco
Tuinier$^{2}$}
\address{$^{1}$ Soft Condensed Matter,
  Debye Institute, Utrecht University, Princetonplein 5,
  3584 CC Utrecht, The Netherlands\\
  $^{2}$
Forschungszentrum J\"ulich, Institut f\"ur Festk\"orperforschung,
52425 J\"ulich, Germany. }
\date{\today}

\noindent

\begin{abstract}
We study the stability of mixtures of highly screened repulsive
charged spheres and non-adsorbing ideal polymer chains in a common
solvent  using free volume theory.
The effective interaction
between charged colloids in an aqueous salt solution is described
by a screened-Coulomb pair potential, which supplements the pure
hard-sphere interaction. 
The ideal polymer chains are treated as spheres that are excluded
from the colloids by a hard-core interaction, whereas the interaction between two ideal chains is set to zero.
In addition, we investigate the phase behaviour of charged
colloid-polymer mixtures in computer simulations, using the
two-body (Asakura-Oosawa pair potential) approximation to the
effective one-component Hamiltonian of the charged colloids. Both
our results obtained from simulations and from  free volume theory
show similar trends. We find that the screened-Coulomb repulsion
counteracts the effect of the \textit{effective} polymer-mediated
attraction. For mixtures of small polymers and relatively large
charged colloidal spheres, the fluid-crystal transition shifts to
significantly larger polymer concentrations with increasing range
of the screened-Coulomb repulsion. For relatively large polymers,
the effect of the screened-Coulomb repulsion is weaker. The
resulting fluid-fluid binodal is only slightly shifted towards
larger polymer concentrations upon increasing the range of the
screened-Coulomb repulsion. In conclusion, our results show that
the miscibility of dispersions containing charged colloids and
neutral non-adsorbing polymers increases, upon increasing the
range of the screened-Coulomb repulsion, or upon lowering the salt
concentration,
 especially when the polymers are small compared to the colloids.
\end{abstract}

\section{Introduction}

\label{intro}
Adding non-adsorbing polymers to colloidal dispersions allows modifying the
 range and strength of attraction of the effective interactions
between the colloidal particles. Adjusting the range of the
attraction enables manipulating the topology of the phase diagram
of a colloid-polymer mixture \cite{Sperry1981, Gast1983,
Lekkerkerker1992, Ilett1995}. Both the nature of the demixed
phases
 as well as the colloid and polymer concentrations at which demixing takes
place depend on the range of attraction \cite{Poon2002}.
Industrially, it is relevant to understand the phase
behaviour of  colloid-polymer mixtures because colloidal
particles and polymer chains are often jointly present in various
products, such as food dispersions
\cite{Grinberg1997,Doublier2000}. Theoretically, the focus has
been mainly on hard spheres dispersed in
 polymer solutions. This is due to the fact that the hard-sphere dispersion
 is relatively simple and well-understood.
Moreover, hard sphere-like colloids can actually be synthesised
chemically.  Examples are, for instance, stearyl-silica colloids
\cite{VrijHeld1980} or PMMA-PHSA particles \cite{PuseyMeg1986}.

In practice, many stable dispersions containing spherical colloids
consist of particles that are not 'pseudo-hard' but which can be
characterised by a pair potential containing an additional soft
repulsive tail. An example is a stable dispersion of charged
colloids in an aqueous salt solution for which the interactions
are described by the Derjaguin-Landau-Verwey-Overbeek (DLVO)
theory \cite{VerweyOverbeek1948}. This theory predicts that the
effective pair interaction between charged colloids consists of a
hard-core repulsion due to the finite size of the colloids and a
screened-Coulomb (Yukawa) repulsion with the screening length
given by the Debye length $\kappa^{-1}$ of the solvent. The
screening length $\kappa^{-1}$ defines the thickness of the double
layer of opposite charge surrounding each colloid. The range
$\kappa^{-1}$ of the screened-Coulomb repulsion is a function of
the salt concentration of the solvent, the dielectric constant and
the temperature. Adjusting the salt concentration may influence
the stability of a dispersion of charged colloids mixed with a
neutral depletion agent in a common aqueous salt solution
\cite{Grinberg1997,Hebert1963,Finet2001,Casselyn2001}. Hebert
\cite{Hebert1963} studied the precipitation of the charged,
rod-like
 tobacco mosaic virus (TMV) by adding the neutral polymer polyethylene
 glycol (PEG).
At similar PEG concentrations, precipitation of TMV was enhanced
by adding salt. Patel and Russel \cite{Patel1989} studied the
phase behaviour of mixtures of charged polystyrene latex colloids
and dextran as (neutral) polymer chains and reported a significant
shift towards higher polymer concentrations of the fluid-fluid
binodal curve as compared
 to predictions for neutral polymer chains mixed with hard spheres.
Grinberg and Tolstoguzov \cite{Grinberg1997} presented generalised phase
diagrams of
proteins mixed with neutral non-adsorbing polysaccharides in aqueous
salt solutions.
The miscibility or compatibility was shown to increase when the ionic
strength of the
 solvent was lowered. The compatibility especially increased below 0.5 $M$.
Finet and Tardieu \cite{Finet2001} studied the stability of
solutions of the lens protein
crystalline. Adding an excess of salt to this system does not destabilise the
protein
dispersion. Hence, it follows that the effective attractions between the
proteins
are absent or are very weak in the case of screened charges. Adding PEG
 however induces significant attractions \cite{Finet2001}, and results in a shift
 of the
liquid-liquid phase transition to higher temperatures \cite{Annunziata2002}.
Adding excess salt and PEG induces instant phase separation
\cite{Finet2001}. A similar synergetic effect of salt and PEG was
found in aqueous solutions of (spherical) brome mosaic virus
particles \cite{Casselyn2001}. In conclusion, the trend found in
 experimental studies on mixtures of charged 'colloids' plus neutral polymers
 is that the miscibility is increased upon decreasing the salt concentration,
i.e., increasing the range of the screened-Coulomb repulsion.

In the light of these findings it is important to study
theoretically mixtures of colloids with a screened-Coulomb
repulsion mixed with neutral polymer chains and to investigate
whether the trend found in many experimental studies is recovered.
The amount of theoretical work performed so far is rather limited.
Ferreira \textit{et al}. \cite{Ferreira2000} made a PRISM analysis
up to the level of the pair interaction and computed gas-liquid
spinodal curves from the effective colloid-colloid structure
factor. Denton and Schmidt \cite{Denton2005} proposed a simple
theory yielding the gas-liquid binodal curve. The fluid-solid
coexistence curves were
 not considered and none of these theories
were tested against computer simulations. Here we study the effect
of a screened-Coulomb interaction  on the total effective
colloid-colloid interaction and on the resulting gas-liquid and
fluid-solid phase transitions in a charged colloid dispersion with
added non-adsorbing polymers. We demonstrate the
 fluid-solid coexistence is especially sensitive to the screened-Coulomb
 repulsion. The outline is as follows.
First, in section \ref{model} we calculate the total effective
pair potential
 by juxtaposing the depletion and the DLVO (screened-Coulomb
repulsion) contributions. In section \ref{phaseYuk}, a simple
Ansatz for  calculating  the phase behaviour of the charged
colloid dispersion is addressed, followed by an analysis of the
effect of non-adsorbing polymer using free volume theory in
section \ref{freevoltheory}. The theoretical results are compared
with Monte Carlo simulation results in section \ref{simulations}.
This communication will be summarised with some concluding
remarks.

\section{The model}

\label{model}

\begin{figure}[ht]
 \begin{center}
 \begin{tabular}{ccc}
  \includegraphics[width=5.5cm]{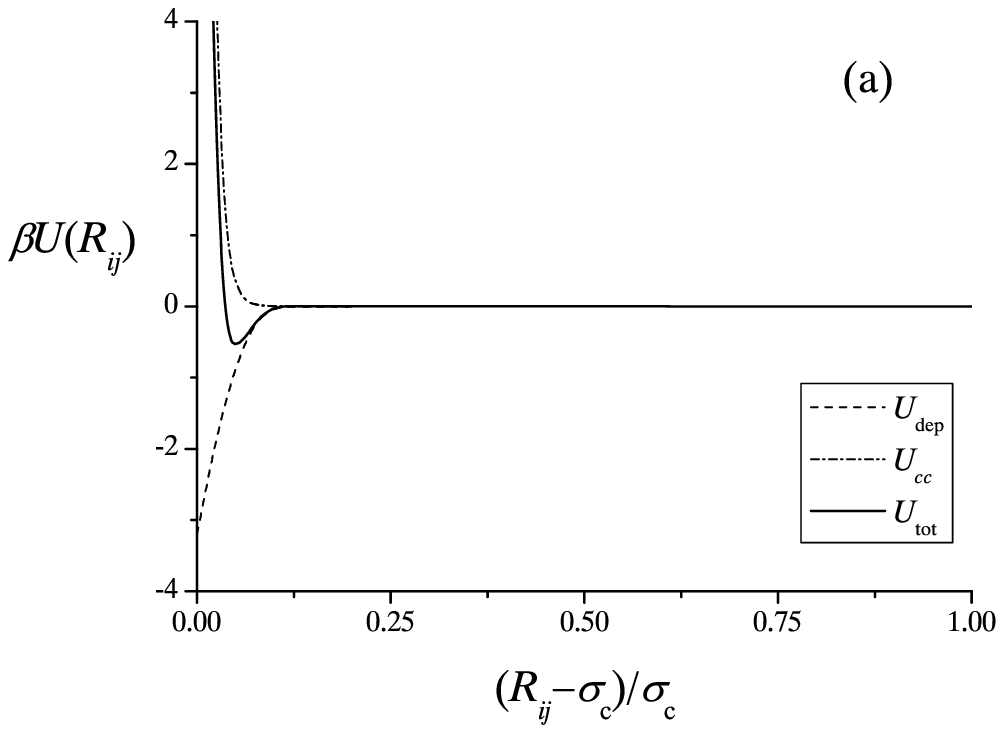}&
  \includegraphics[width=5.5cm]{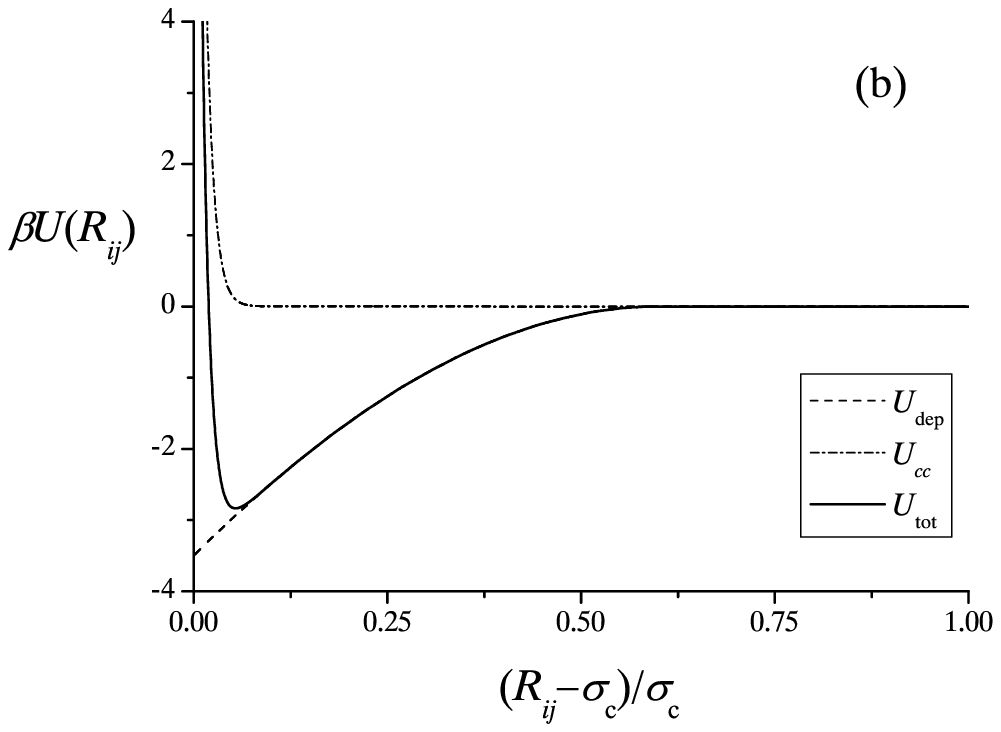}&
  \includegraphics[width=5.5cm]{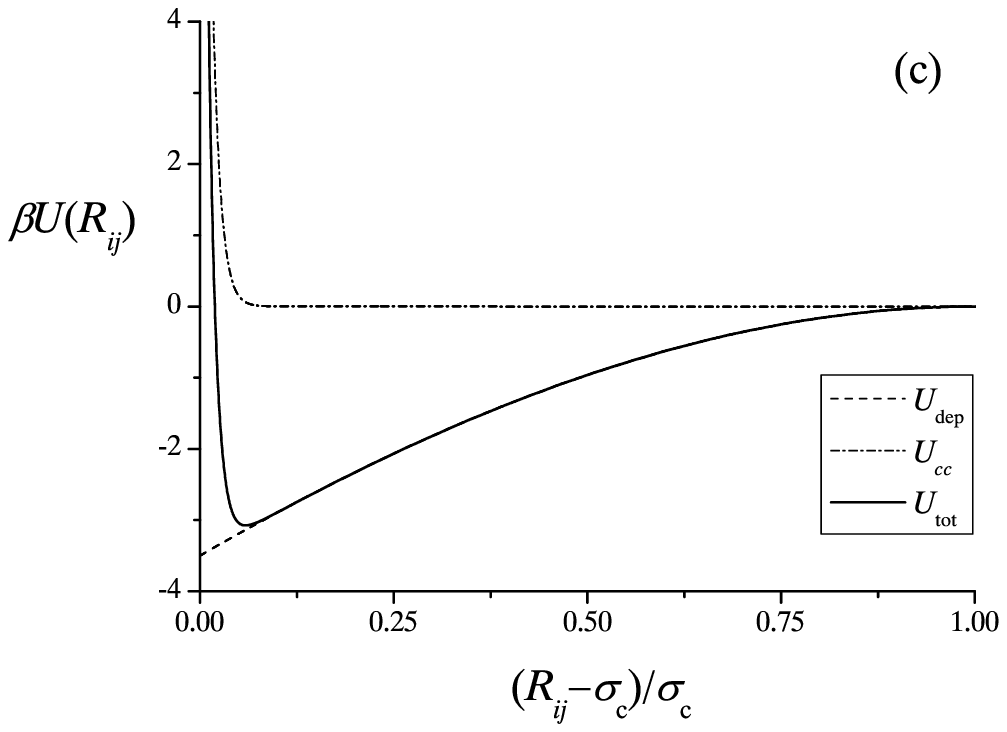}\\
 \end{tabular}
  \end{center}
 \caption{Pair interaction between two colloidal spheres with hard-core diameter $\sigma_c$ interacting with a
Yukawa pair potential $U_{cc}$ (equation \ref{Yukint}) and an
effective
 depletion interaction $U_{\mbox{\footnotesize dep}}$ (equation
\ref{AO}) due to the presence of non-adsorbing polymer coils with
an effective diameter $\sigma_p$. The total effective interaction
$U_{\mbox{\footnotesize{tot}}}$ (equation \ref{totint}) is denoted
by the full curve. The Yukawa repulsion is characterised by $\beta
\epsilon = 20$ and $\kappa \sigma_{c} =100$. The depletion
interaction are: a) for a size ratio $q=\sigma_{p}/\sigma_{c}=0.1$
and relative polymer concentration $\phi^{r}_{p}$=0.2, b) $q=0.6$
and $\phi^{r}_{p}$=1, and c) $q=1$ and $\phi^{r}_{p}$=1.4.}
\label{fig1}
\end{figure}

We consider a suspension of charged colloidal spheres immersed
together with non-adsorbing polymer in a common solvent. As the
differences in length- and time-scales between the solvent
molecules and the colloids and polymers are large, we can regard
the solvent to be an inert continuum. 
Charged-stabilized colloidal suspensions can be described by a
mixture of $N_c$ colloidal particles with charge $-Ze$, with $e$
the elementary charge, $Z N_c$ counterions with charge $+e$ and by
pairs of salt ions of charge $\pm e$. It is convenient to consider
the system in equilibrium with a salt reservoir with density
$\rho^r_s$. However, computer simulations of this model are
prohibited by slow equilibration, due to a large number of small
counterions and salt pairs compared to the number of colloidal
particles. A more coarse-grained model can be obtained by formally
integrating out  the degrees of freedom of the co- and counterions
in the partition function \cite{Roij1999}. The effective pair
interaction between the charged colloidal spheres with hard-core
diameter $\sigma_c$, reads
\begin{equation}
U_{cc}(R_{ij}) = \left \{\begin{array}{ll}
\infty & \mbox{for}\hspace{0.5cm} R_{ij}<\sigma_c \\
                \epsilon \left (\frac {\exp (- \kappa
               \sigma_c(R_{ij}/\sigma_c-1))}{R_{ij}/\sigma_c}\right)&
                 \mbox{otherwise} \end{array} \right. ,
                \label{Yukint}
\end{equation}
where $R_{ij}= |{\bf R}_i-{\bf R}_j|$ and ${\bf R}_i$ are the
positions of the centres of the colloids. The range of the
repulsive tail is set by the inverse Debye screening length
$\kappa\sigma_c= \sqrt{8 \pi l_B \sigma_c^2 \rho^r_s}$, which is
related to the salt concentration $\rho^r_s$ in the reservoir, and
to the Bjerrum length $l_B=e^2/\epsilon_s kT$ with $k$ Boltzmann's
constant, $T$ the temperature, and $\epsilon_s$ the dielectric
constant of the solvent. The strength of the repulsion of the
repulsive Yukawa interaction is determined by the parameter
$\epsilon = (Z/(1+\kappa\sigma_c/2))^2 l_B/\sigma_c$. We neglected
the higher order, effective many-body terms which are
insignificant in the high salt concentration regime that is
considered  in this work \cite{Roij1999}. In this approximation,
the effective potential (\ref{Yukint}) is the usual DLVO potential
\cite{VerweyOverbeek1948,Pier1983,Robbins1988,bookRuss1989,Loewen1993,Stevens1993,
Dijkstra2001,Russ}. In figure \ref{fig1} an example of a typical
Yukawa repulsion is given by the dot-dashed curves for $\beta
\epsilon = 20$ and $\kappa \sigma_{c} =100$. Here $\beta = 1/kT$.

Different procedures
\cite{Graf1998,Denton2000,Warren2000,Belloni2000,Dijkstra1998-I}
exist that give the same functional form of the effective
potential (\ref{Yukint}), but with a density dependent $\tilde
\kappa\sigma_c=\sqrt{( 4 \pi l_B \sigma_c^2(Z \rho_c + 2\rho_s)
)}$, where $\rho_c$ and $\rho_s$ are, respectively, the densities
of colloidal particles and added salt ion pairs, in the system.
The density $\rho_s$ depends on the colloidal density and is
smaller then the reservoir density due to salt exclusion or Donnan
equilibrium \cite{Donnan1911,Donnan1911-I,Donnan1924}. For the
parameters used in this work,  $\beta \epsilon = 20$ and $\kappa
\sigma_{c} =50$, the effective $\tilde \kappa$ differ from the
reservoir $\kappa\sigma_c= \sqrt{8 \pi l_B \sigma_c^2 \rho^r_s}$ by less then 1$\%$ for a
colloid density near close-packing $\rho_s=1.4 \sigma_c^{-3}$ when
one applies Donnan equilibrium. 
Hence salt partitioning is hardly perceptible for the salt conditions considered here.

A dilute solution of polymers in a theta-solvent can be represented by
(non-interacting) ideal polymers. In a theta-solvent, the
attraction between the polymer segments compensates their excluded
volume: the chains do not feel one another. The non-interacting
polymer coils are excluded from the colloids to a centre-of-mass
distance $(\sigma_c+\sigma_p)/2$ \cite{Asakura1958}. The chains
can be interpreted as freely overlapping spheres
\cite{Vrij1976,DeHek1981}. The effective diameter of the polymer
coil $\sigma_p$ is twice the depletion thickness around the
colloidal particle. The pair-wise potentials in this simple model
read \cite{AOnote}
\begin{eqnarray}
U_{cp}({\bf R}_i-{\bf r}_j) & = & \left \{\begin{array}{ll} \infty
& \mbox{for} \hspace{5mm}
|{\bf R}_i-{\bf r}_j|<(\sigma_{c}\;+\;\sigma_{p})/2 \\
             0    & \mbox{otherwise} \end{array} \right.\\
U_{pp}(r_{ij}) &=& 0
\end{eqnarray}
Where ${\bf R}_i$ and ${\bf r}_j$ are the positions of the centres
of the colloids and the polymer coils, respectively, and  $r_{ij} = |{\bf r}_i-{\bf r}_j|$.
We mention that the so-called Asakura-Oosawa-Vrij (AOV) model
\cite{Asakura1958, Vrij1976,DeHek1981}, which is a simple
idealised model for colloidal hard spheres and ideal non-adsorbing
polymer, is recovered by setting $\epsilon$ to zero in the effective potential  (\ref{Yukint}).

However, even within the context of this highly simplified model
it is often more convenient to adopt a more coarse-grained view of
the binary mixture by ignoring the degrees of freedom of the
polymer coils and using polymer-mediated effective interactions
between the charged spheres. Below
 we discuss the resulting effective interactions between
the charged spheres due to the pair interactions of our model.

By integrating out the degrees of freedom of the polymer coils, we
can map the binary mixture of colloids and polymers onto an
effective one-component system interacting with an effective
one-component Hamiltonian
\cite{Dijkstra1998,Dijkstra1999L,Dijkstra1999E,Dijkstra1999}. This
effective Hamiltonian consists of zero-body,
 one-body, two-body, and higher-body terms. In this work, we perform
 Monte-Carlo simulations for the effective Hamiltonian truncated at
 the pairwise term, and we determine the phase behaviour of the effective
 one-component system. We first derive an exact expression for the
polymer-mediated effective pair potential or depletion potential.
In general the effective pair potential  can be calculated using
the generalised Gibbs adsorption equation for two colloids
separated by a distance $R_{ij}$ in a sea of polymer at fixed
chemical potential $\mu_p$, i.e., the system of colloids plus
polymers is in thermodynamic equilibrium with a reservoir with
only polymer chains (in a background solvent) at chemical
potential $\mu_p$
\cite{DGHall1972,Ash1973,Tuinier2000,Tuinier2003}

\begin{equation} \label{Gibbsadseq}
\beta U_{\mbox{\footnotesize
dep}}({R}_{ij},\mu_p)\;=\;-\int_{-\infty}^{\mu_p} \beta
d\mu_{p}{'}
\left(\Gamma({R}_{ij},\mu_p)\;-\;\Gamma({R}_{ij}=\infty,\mu_p)
\right ).
\end{equation}
The chemical potential is related to the fugacity of the polymer
chains $z_p$ through $\beta d\mu_p= d \ln z_p$. The quantity
$\Gamma({R}_{ij},\mu_p)$ is the adsorption or excess amount of
polymer chains when the colloidal spheres are separated
by a distance ${R}_{ij}$.

The depletion potential of the present model is identical to the
depletion potential of the AOV model as the pair potentials
$U_{cp}$ and $U_{pp}$ are the same in both models. Since the
polymer chains do not interact, $U_{pp}=0$, we can replace $z_p$
by $\rho_p^r$, where $\rho_p^r$ is the polymer density in the
corresponding polymer reservoir.
 Further within the model the depletion layers have a fixed width of
 $\delta \equiv \sigma_p /2$ being close to the radius of gyration of a polymer chain
 \cite{Eisenriegler1983,Tuinier2000,Fleer2003}, so the excess adsorbed amounts are directly related to the overlap
volume $V_{overlap}({R}_{ij})$ of depletion layers:
\begin{eqnarray*}
\label{eqadsamounta} \Gamma({R}_{ij},\mu_p)\;-\;\Gamma({R}_{ij}=\infty,\mu_p)
\;=\;\rho_p^r V_{overlap} ({R}_{ij})
\end{eqnarray*}
A geometrical calculation for
$\sigma_c<{R}_{ij}<\sigma_c+\sigma_p$,
 the range where the depletion layers overlap, yields an analytic
 expression for $V_{overlap}(R_{ij})$ and the depletion potential
 $U_{\mbox{\footnotesize dep}}$ reads for our model \cite{Asakura1958,Vrij1976,DeHek1981,Dijkstra1999}
\begin{eqnarray} \beta U_{\mbox{\footnotesize
dep}}({R}_{ij},\mu_p)&=& -\frac{\pi \sigma_p^3 \rho_p^r}{6}
\frac{(1+q)^3}{q^3}\left \lbrack 1- \frac{3 R_{ij}}{2
(1+q)\sigma_c} \;+\;
\frac{R_{ij}^3}{2(1+q)^3\sigma_c^3} \right \rbrack \nonumber \\
& &
\hspace{10mm}\mbox{for}\hspace{3mm} \sigma_c<R_{ij}<  \sigma_{c}+\sigma_p \label{AO}\\
& = & 0 \hspace{10mm}\mbox{for}\hspace{3mm} R_{ij}>
\sigma_{c}+\sigma_p \nonumber
\end{eqnarray}
where we define the size ratio $q=\sigma_p/\sigma_c$. This
Asakura-Oosawa pair potential describes an effective attraction
 whose depth increases linearly with the polymer density in the
corresponding reservoir $\rho_p^r$. For convenience, we
define the relative polymer concentration $\phi_{p}^r \equiv
 \pi \sigma_{p}^3 \rho_{p}^r/6$. Hence, $\phi_{p}^r=1$
defines the overlap concentration of the polymer solution. The
range of the potential is given by $\sigma_p$. The depletion
interaction $U_{\mbox{\footnotesize{dep}}}(R_{ij})$ is plotted as
the dashed curves for $q=0.1$ and $\phi_{p}^r$=0.2 in figure \ref{fig1}a,
for $q=0.6$ and $\phi_{p}^r=1$ in figure \ref{fig1}b and for $q=1$ and
$\phi_{p}^r=1.4$ in figure \ref{fig1}c. By adjusting $q$ and $\phi_{p}^r$
one can manipulate the range and strength of the depletion
interaction. We chose $\phi_{p}^r$ such that $\beta
U_{\mbox{\footnotesize{dep}}} (\sigma_c)=-3.5$ for every value of
$q$.

It is important to note that for sufficiently large polymer coils,
effective three- and higher-body interactions can not be
neglected. More precisely, we expect an increasing number of
higher-body interactions to become non-zero when $q$ increases. It
was shown \cite{Dijkstra2002} that the many-body character of the
polymer-mediated effective interactions between the colloids
yields a bulk phase diagram that differs substantially from those
found for pair-wise simple fluids. For size ratios $q<0.1547$,
three- and many-body interactions are identical to zero and the
mapping of the binary mixture onto the effective one-component
Hamiltonian based on pair-wise additive effective potentials is
exact.

Using the two-body approximation to the effective Hamiltonian, the
total effective pair potential $U_{\mbox{\footnotesize{tot}}}$ reads

\begin{equation} U_{\mbox{\footnotesize{tot}}} (R_{ij})\;=\;U_{cc}
(R_{ij})\;+\;U_{\mbox{\footnotesize{dep}}} (R_{ij}).
\label{totint}
\end{equation}
Examples of $U_{\mbox{\footnotesize{tot}}}(R_{ij})$ are plotted in
figure \ref{fig1} as the full curves, which are the sums of equations
(\ref{Yukint}) and  (\ref{AO}), denoted by the dot-dashed and
dashed lines, respectively.
 For $q=0.1$ (figure \ref{fig1}a) there is a significant effect of the repulsive tail
 on the effective interaction as compared to the pure depletion contribution
 $U_{\mbox{\footnotesize{dep}}} (R_{ij})$.
There is, however, still some attraction in the
$U_{\mbox{\footnotesize{tot}}}(R_{ij})$ curve between the charged
repulsive spheres though it is significantly reduced as compared
to the pure $U_{\mbox{\footnotesize{dep}}} (R_{ij})$ result. In
figures \ref{fig1}b and \ref{fig1}c the main parts of the pair interaction curves
$U_{\mbox{\footnotesize{tot}}}(R_{ij})$ are identical to the pure
depletion part $U_{\mbox{\footnotesize{dep}}}(R_{ij})$ for $q=0.6$
and $q=1$. Only the attraction at short interparticle distances is
affected. The results plotted in figure \ref{fig1} are qualitatively
similar to the PRISM results of Ferreira \textit{et al}.
\cite{Ferreira2000} (see their figure 5).

\section{Phase behaviour of Charged colloids}

\label{phaseYuk}

In this section we propose a simple description for the phase
behaviour of  charged colloidal suspensions. We consider the
screened-Coulomb repulsion as a perturbation of the hard sphere
interaction which is only valid for highly screened colloidal
suspensions. The effective volume fraction $\eta_{e}$ of the
charged repulsive spheres is approximated as
\begin{equation} \label{eff vol frac}
\eta_{e}\;=\;\left(\frac
{\sigma_{e}}{\sigma_{c}}\right)^3\;\eta_c\;=\;m \; \eta_c
\end{equation}
where $\eta_c \equiv \pi \sigma_c^3 \rho_c/6$ with $\rho_c$ the
colloid number density and $\sigma_{e}$ the effective diameter of
the spheres defined by the Barker-Henderson relation
\cite{BarkerHenderson1967}:
\begin{equation} \label{Barkereq}
\sigma_{e}\;=\;\sigma_{c}\;+\;\int_{\sigma_{c}}^{\infty}\left(1\;-\;\exp
[-\beta U_{cc}(r)]\right) dr
\end{equation}
which is a useful way to generalise pair potentials of various
shapes \cite{LouisRoth2001}, in particular, perturbations from
hard-sphere behaviour (see  \cite{HansenMcDonald1986}). Instead of
$U_{cc}$ defined in equation (\ref{Yukint}) one may use any other
form for a (soft) repulsion. The physical effects are contained in
the value for $m$. Since we assume that the collection of charged
spheres behave similarly as a collection of pure hard spheres plus
a small perturbation, we may use the Carnahan-Starling (CS)
expression \cite{Carnahan1969} for the Helmholtz free energy to
describe the thermodynamic properties of the fluid of charged
spheres:
\begin{equation} \label{CarnStarfeq}
mf_{c}^{\mbox{\footnotesize{fluid}}}(\eta_c,T)\;=\;\eta_{e} \ln
\eta_{e}\; +\;
\frac{4\eta_{e}^2\;-\;3\eta_{e}^3}{(1\;-\;\eta_{e})^2}\;-\;\eta_{e}
+\eta_e\ln \frac{6\Lambda_c^3}{\pi\sigma_c^3}
\end{equation}
where $\Lambda_c$ is the thermal wavelength of the colloids and
where the hard sphere volume fraction $\eta_{c}$ in the classical
CS expression is replaced by the effective volume fraction
$\eta_{e}$, defined in equations (\ref{eff vol frac}) and
(\ref{Barkereq}). In equation (\ref{CarnStarfeq}) we use (as in
\cite{Aarts2002}) the normalised Helmholtz free energy $f_{c}$,
defined as $\beta F_c v_c/ V$, where $F_c$ is the Helmholtz free
energy, and $v_c=\pi\sigma_c^3/6$ is the volume of a single
colloid. We note that there are more advanced theoretical methods
for Yukawa fluids \cite{cochran} that describe the simulation
results for longer-ranged repulsions.

The equation of state of the face-centred-cubic (fcc) crystal
phase of pure hard spheres is described  accurately by the
expression proposed by Hall \cite{CKHall1972}. Likewise, the
equation of state for the fcc crystal phase of the charged
 spheres reads
\begin{equation}\label{Hallfeq}
mf_{c}^{\mbox{\footnotesize{crystal}}}(\eta_c,T)\;=\; \eta_{e}
\left ( 2.1306\; +\; 3 \ln \left
[\frac{\eta_{e}}{1-\eta_{e}/\eta_{cp}} \right]\right) +\eta_e\ln
\frac{6\Lambda_c^3}{\pi\sigma_c^3}
\end{equation}
containing the volume fraction at close packing
$\eta_{cp}=\pi\sqrt{2}/6 \approx 0.74 $. The quantity $2.1306$ is
derived from computer simulation results \cite{Frenkel1984}.

\begin{figure}[ht]
 \begin{center}
 \includegraphics{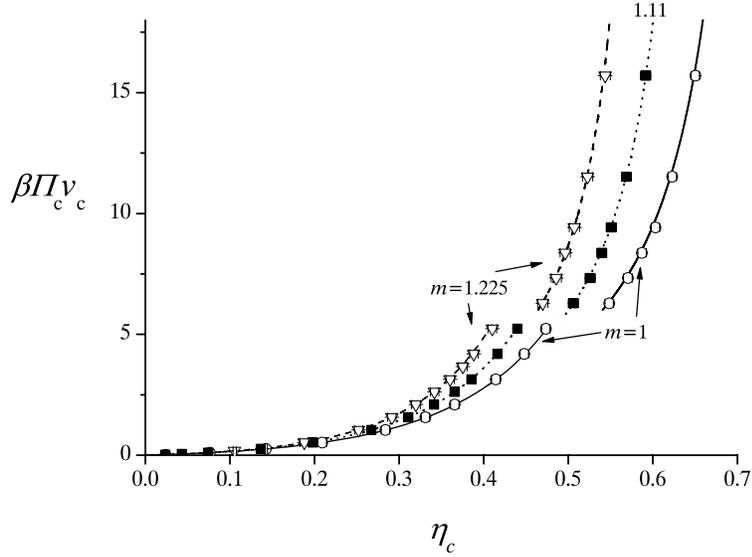}
  \end{center}
\caption{Pressure of a dispersion of charged colloidal spheres
interacting with $U_{cc}(R_{ij})$ (see equation (\ref{Yukint}))
for $\beta \epsilon = 20$ and $1/\kappa\sigma_c=0$ (open circles,
full curves), 0.01 (filled squares, dotted curves), and 0.02 (open
triangles, dashed curves), corresponding  to $m=1, 1.110$, and
1.225, respectively, in equation (\ref{eff vol frac}). The symbols
denote the simulation results, while the curves denote the
theoretical predictions of equations (\ref{CarnStarpresseq})
(lower set of pressures) and (\ref{Hallpresseq}) (upper set).}
\label{fig2}
\end{figure}

Let us first check the accuracy of equations (\ref{CarnStarfeq})
and (\ref{Hallfeq}) on the level of the resulting osmotic pressure
$\Pi_{c}$ that follows from $\beta \Pi_{c} v_c=\eta_{c} \partial
{f_{c}} / {\partial \eta_{c}}-f_{c}$. The osmotic pressure for the
fluid phase reads:
\begin{equation} \label{CarnStarpresseq}
m \beta \Pi_{c}^{\mbox{\footnotesize{fluid}}} v_{c}(\eta_c,T)
\;=\;\frac{\eta_{e}\;+\;\eta_{e}^2\;+\;\eta_{e}^3\;-\;\eta_{e}^4}
{{(1\;-\;\eta_{e})}^3}
\end{equation}
and for the fcc crystal, the pressure is:
\begin{equation} \label{Hallpresseq}
m \beta \Pi_{c}^{\mbox{\footnotesize{crystal}}} v_{c}(\eta_c,T)
\;=\;\frac{3 \eta_{e}} {1\;-\;\eta_{e}/\eta_{cp}}
\end{equation}
Results using equations (\ref{CarnStarpresseq}) and
(\ref{Hallpresseq}) with $U_{cc}(R_{ij})$ given by equation
(\ref{Yukint}) are plotted in figure \ref{fig2} (curves) and are compared
with computer simulation
 data (symbols), for $\beta \epsilon = 20$ and $1/\kappa \sigma_{c}$ equals 0
 (full curves, open circles), 0.01 (dotted curve, filled squares) and 0.02 (dashed curve, open triangles), corresponding to
$m=1$, $1.110$ and $1.225$, respectively. The results for
pressures below $\beta \Pi_{c} v_c=6.2$ correspond to the
colloidal fluid phase, while the results for larger pressures
correspond to the  fcc crystal. Figure \ref{fig2} shows that the pressure
increases upon increasing the range of the soft repulsion.
 The simulation results are well described by equation
 (\ref{CarnStarpresseq})
(see the results for $\beta \Pi_c v_c <6.2$.) for the fluid phase
 for $m=1$, $m=1.110$ and $m=1.225$.   In addition, the results  for the solid phase using
equation (\ref{Hallpresseq}) ($\beta \Pi_c v_{c}>6.2$) agree well
with  the simulations  for $m=1$, $m=1.110$ and $m=1.225$. Our
results  demonstrate that for sufficiently short-ranged soft
repulsions, the screened-Coulomb interaction can be treated as a
perturbation of the hard sphere potential using the
Barker-Henderson relation (\ref{Barkereq}).

\begin{figure}[ht]
 \begin{center}
  \includegraphics[width=7cm]{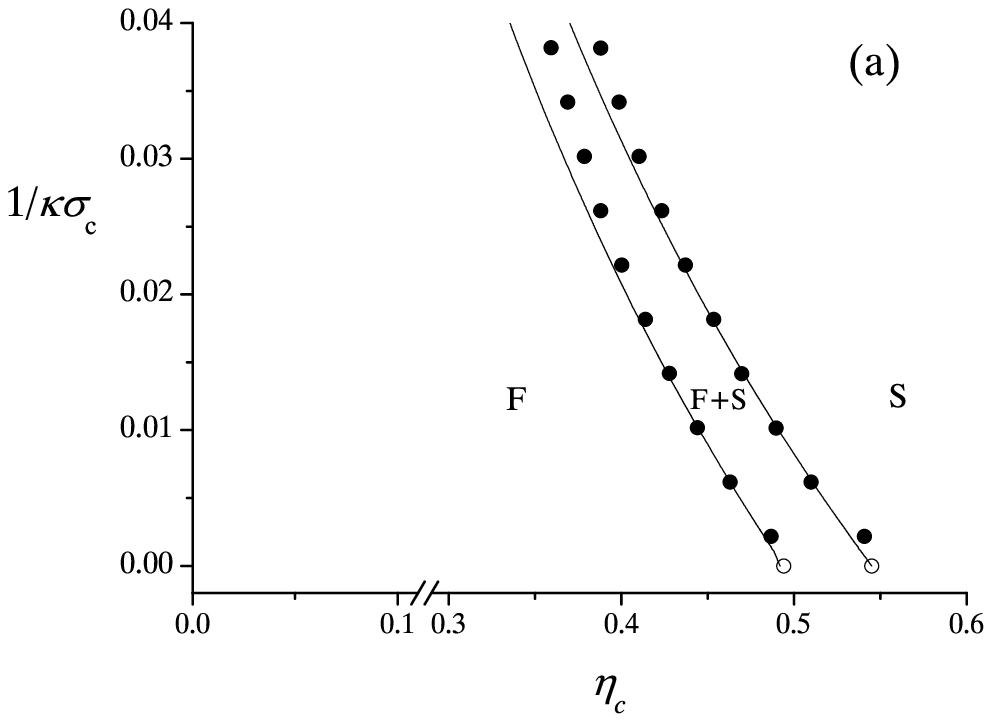}
  \includegraphics[width=7cm]{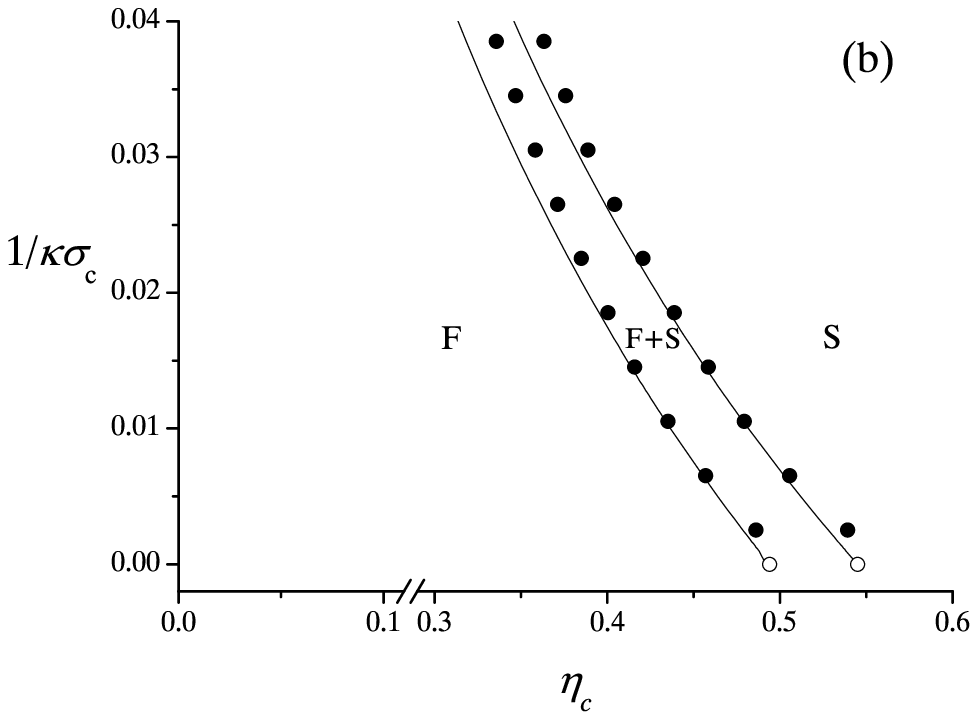}
 \end{center}
 \caption{Fluid-solid (fcc) transition of charged colloidal spheres interacting with a
 hard-core repulsive Yukawa potential (\ref{Yukint})
 with a) $\beta \epsilon = 20$ and b) $\beta \epsilon = 39$.
Open symbols denote simulation results of pure hard spheres taken
from Ref. \cite{Hoover1968}. Filled circles denote the simulation
results of Hynninen and Dijkstra \cite{Hynninen2003}. Full curves
correspond to the theoretical predictions as described in section
\ref{phaseYuk}.}
\label{fig3}
\end{figure}

The fluid-solid transition, first established for pure hard
spheres by Alder and Wainwright \cite{Alder1957} and Wood and
Jacobson \cite{Wood1957}, can now be studied as a function of the
softness of the repulsive tail. We determine the densities of the
coexisting phases  by equating the osmotic pressures and the
chemical potentials $\beta \mu=\partial {f_{c}} / {\partial
\eta_{c}}$ using equations (\ref{CarnStarfeq}) and
(\ref{Hallfeq}). It becomes evident that the Ansatz of equations
 (\ref{eff vol frac}), (\ref{Barkereq}), (\ref{CarnStarfeq})
and (\ref{Hallfeq}) is expected to be useful only for short-ranged
soft repulsions, so for $\sigma_{e} \simeq \sigma_{c}$  or
$1/\kappa \sigma_{c} \to 0$. In order to test whether this
approach is valid for relatively short-ranged soft repulsions, we
compare the predicted fluid-solid transitions with computer
simulation data for small values of $1 / \kappa \sigma_{c}$. In
figure \ref{fig3}, the fluid-solid binodals calculated for $1 / \kappa
\sigma_{c}<0.04$ are plotted as the full curves. In figure \ref{fig3}a, the
$\beta \epsilon=20$ case is considered. The dots are computer
simulation results by Hynninen and Dijkstra \cite{Hynninen2003}.
Open symbols are the pure hard sphere computer simulation results
of Hoover and Ree \cite{Hoover1968}. For $1 / \kappa
\sigma_{c}<0.02$ the agreement is excellent though for larger
Debye lengths deviations are found. This deviation is not
surprising since the simple theoretical method is based on a
perturbation from the hard sphere system and is, hence, expected
to be only accurate for very small values of $1 / \kappa
\sigma_{c}$. Still, this simple theory for charged colloidal
sphere dispersions suffices our purpose of studying the stability
of charged colloid-polymer mixtures in the regime of $1 / \kappa
\sigma_{c}<0.02$. To be more specific, this means that our Ansatz
describes well the case of colloidal particles with a diameter of,
say, $\sigma_{c}=100$ nm with a  Debye length
 smaller than about 2 nm or, equivalently, an ionic strength $>
0.02 M$ (in the case of monovalent ions in water at room
temperature). Actually, in nature, the ionic strength in aqueous
salt concentrations is usually above $0.02 M$, so for large
spheres we still capture the biologically relevant ionic strength
regime. For globular proteins we can only compare with
experimental data at significant salt concentrations but we can
still make estimates of the main trends.

In figure \ref{fig3}b, we compare our theoretical results (full curves)
with computer simulation data (crosses) for $\beta \epsilon=39$.
Again, our Ansatz is in fair agreement with the computer
simulation results (data points) \cite{Hynninen2003}. Hence we
validated our approach to describe the fluid and fcc crystal
equations of state and the fluid-solid transition of a charged
sphere dispersion for $1 / \kappa \sigma_{c}<0.02$. This provides
a base for studying the effect of adding non-adsorbing polymer to
such a suspension.

\section{Phase behaviour of Charged colloid-polymer mixtures}

\label{phasepolYukcoll}
\subsection{Free volume theory}
\label{freevoltheory}

\begin{figure}[ht]
 \begin{center}
  \includegraphics{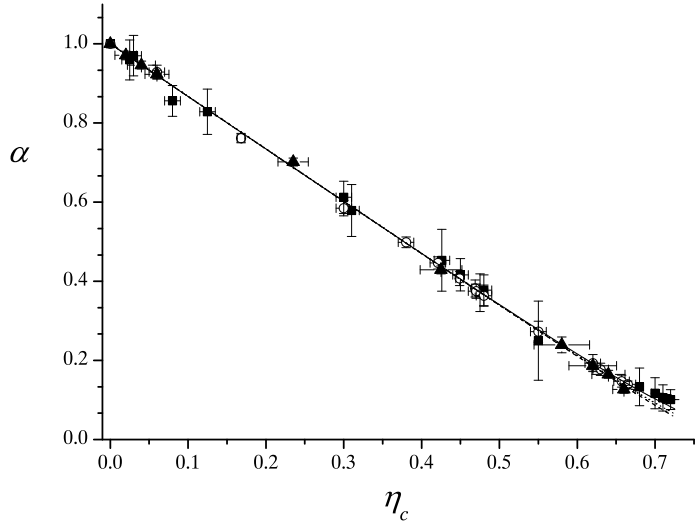}
   \end{center}
 \caption{Free volume fraction $\alpha \equiv \langle V_{free}
 \rangle_{z_p=0}/V$ for a mixture of charged colloids and ideal polymer with
 size ratio $q=\sigma_p/\sigma_c=0.1$ as a function of hard-core volume
 fraction $\eta_{c}$. The screened-Coulomb repulsion is characterised by
 $\beta \epsilon = 20$ and various values of $\kappa \sigma_{c}$.
 Full, dotted and dashed curves represent equation
 (\ref{eqalfa}) for  $\kappa \sigma_{c}=\infty \;(m=1)$, $100 \;(m=1.110)$
 and $80 \;(m=1.138)$, respectively. Note that the differences in the
 theoretical curves are only noticeable at high $\eta_c$.
 The symbols with errorbars denote the  Monte Carlo
 simulation results for $\langle \rho_p\rangle_{z_p}/\rho_p^r$ using equation (\ref{alpha}) for
 $\kappa \sigma_{c}=\infty$ (closed squares), $\kappa \sigma_{c}=100$ (open
circles) and  $\kappa \sigma_{c}=80$(closed triangles), where we
used $z_p$ at bulk coexistence (see figure \ref{fig11}).}
\label{fig4}
\end{figure}

We study the stability of  a mixture of charged colloidal spheres
and non-adsorbing polymer chains in a common (background) solvent.
A simple approach that successfully describes the stability of
polymer-colloid mixtures is the semi-grand canonical free volume
or osmotic equilibrium theory
\cite{Lekkerkerker1992,Lekkerkerker1990}. In this approach, a
macroscopic volume $V$ at temperature $T$ is considered, that
contains $N_c$ colloids together with polymer chains and solvent,
which are in osmotic equilibrium with a reservoir containing only
solvent and polymer chains. Hence, the system is considered in the
$(N_c,V,z_p,T)$ ensemble, in which the number of colloids $N_c$
and the fugacity of the polymer chains $z_p$ are fixed. Consider
the thermodynamic identity \cite{Dijkstra1999}:
\begin{equation}
\label{eqomegageneral} \beta F(N_c, V, z_p) \;=\;\beta F(N_c,
V, z_p=0) \;+\;\int_{0}^{z_p} d z_p{'}\left (\frac{\partial{\beta
F(N_c, V, z_p{'})}}{\partial{z_p{'}}} \right)_{z_p{'}}
\end{equation}
where we dropped the temperature dependence for convenience.  The
integrand can now be Taylor expanded about $z_p=0$:
\begin{eqnarray}
\label{eqomegageneral1} \beta F(N_c, V, z_p) &=&\beta F(N_c,
V,z_p=0) +z_p\left (\frac{\partial{\beta F(N_c, V, z_p)
}}{\partial{z_p}} \right)_{z_p=0} \nonumber \\
& & \mbox{} + {\cal O}(z_p^2) + \cdots
\end{eqnarray}
where the partial derivative can be written as
\begin{equation}\label{derivative}
\left (\frac{\partial{\beta F(N_c, V, z_p)
}}{\partial{z_p}}\right )_{z_p=0} = \left
(\frac{\partial{F}}{\partial \mu_p}\right
)_{z_p=0}\frac{\partial \beta \mu_p}{\partial z_p}  =
-\frac{\langle N_p \rangle_{z_p=0}}{z_p}.
\end{equation}
The number of polymer chains $\langle N_p \rangle_{z_p=0}$ can be
related by definition to the averaged free volume that is
available for the polymer chains $\langle V_{free}
\rangle_{z_p=0}$ in the system of spheres that is
 undistorted by the addition of polymers:
\begin{equation}
\langle N_p \rangle_{z_p=0}\equiv \rho_p^r \langle V_{free}
\rangle_{z_p=0}. \end{equation} where $\rho_p^r$ is the density of
ideal polymer in the corresponding reservoir. Defining the free
volume fraction $\alpha \equiv \langle V_{free} \rangle_{z_p=0}
/V$, we can rewrite equation (\ref{eqomegageneral}) as:
\begin{eqnarray}
\label{eqomegamodel} \beta F(N_c,V,z_p)\;=\;\beta
F(N_c,V,z_p=0)\;-\; \alpha  \rho_p^r V + {\cal O} (z_p^2) +
\cdots
\end{eqnarray}
The free volume theory \cite{Lekkerkerker1992} retains only the
first-order term, neglecting terms ${\cal O} (z_p^2)$ and higher.
With this assumption the normalised thermodynamic potential
$f\equiv \beta F v_c/V$ can be written as the sum of two
terms:
\begin{equation}
\label{eqomegamodel1} f(N_c, V, z_p) = f_{c}(N_c,V) - \alpha
\phi_p^r q^{-3}
\end{equation}
The first term in equation (\ref{eqomegamodel1}) is the normalised
Helmholtz free energy $f(N_c,V,z_p=0)\equiv f_c(N_c,V)$ of a 'pure colloid' fluid at
a given $\eta_c$, while the second can be interpreted as a
perturbation due to the presence of polymer chains.  Note that
equation (\ref{eqomegageneral})-(\ref{eqomegamodel1})  holds for
any colloid-polymer and polymer-polymer interactions within the
assumptions that are made. However, for interacting polymers, a
different approximation of the free volume theory gives a better
description of the experimental phase diagrams \cite{Aarts2002}.
All information about the interactions between colloid and polymer
is contained in the variation of $\alpha$ with $\eta_c$. For our
model and the AOV model, the free volume fraction $\alpha$ can be
calculated accurately from scaled particle theory
\cite{Lebowitz1965,Lekkerkerker1992} (see \textit{e.g.} Meijer
\cite{Meijer1994} for a comparison with computer simulation
results). Once the coexisting colloid volume fractions are
determined for given $\phi_{p}^{r}$, the actual relative polymer
concentrations can be obtained in the coexisting phases from
$\phi_{p}\equiv \alpha \phi_p^r$.  An expression for the free
volume fraction available for the polymer chains in a mixture of
charged spheres using scaled particle theory reads
\begin{equation}
\label{eqalfa} \alpha\;=\;(1\;-\;\eta_c) \exp(- A\gamma\;
-\;B\gamma^2\;-\;C\zeta\; -\;3C\zeta^2\;-\;3C\zeta^3 ),
\end{equation}
where $\gamma= \eta_c/(1-\eta_c)$, $\zeta=\eta_e/(1-\eta_e)$,
$A\;=\;3q\;+\; 3q^2$, $B\;=\; 9 q^2/2 $, and $C\;=\; q^3$. An
explicit derivation is given in Appendix 1. Equation
(\ref{eqalfa}) reduces to the classical expression of Lekkerkerker
\textit{et al.} \cite{Lekkerkerker1992} for the case
$\eta_{e}=\eta_{c}$ ($m=1$). The free volume available for the
polymer chains is thus mainly a function of the pure hard sphere
volume fraction
 $\eta_{c}$; it is only affected by the screened-Coulomb repulsion at high colloid
volume fractions or large polymer chains. In the derivation of
equation (\ref{eqalfa}) an equal statistical weight is assigned to
all (non-overlapping) hard-sphere configurations, whereas the
weight should involve the polymer-mediated effective interactions
and the screened-Coulomb interactions. We expect our results to be
accurate only if $\sigma_{e}/\sigma_{c}-1<<q$. If the depletion
layers become small compared to $\sigma_{e}-\sigma_{c}$ one
expects hardly any overlap of depletion layers. In figure \ref{fig4} we
compare the result of the free volume fraction $\alpha$ as a
function of the colloid volume fraction $\eta_{c}$ of equation
(\ref{eqalfa}) with Monte Carlo simulation results for $\langle
\rho_p\rangle_{z_p}/\rho_p^r$ using equation (\ref{alpha}) for
$q=0.1$. It is worth mentioning that $\alpha$ is evaluated in the
pure charged colloid system in the free volume theory, i.e.,
$z_p=0$, while $\langle \rho_p\rangle_{z_p}/\rho_p^r$ from
simulations do depend on $z_p$. In the simulations, however, we
use $z_p$ along the bulk binodals as shown in figure \ref{fig1}1 (which
will be discussed later). The technical  details of the
simulations are described in Appendix 2.  The theoretical curves
in figure \ref{fig4} show  only a slight effect of the screened-Coulomb
repulsion on $\alpha$ for $\eta_{c}>0.4$. Within the statistical
error bars,  no significant effect of the screened-Coulomb
repulsion was detected in the simulations on the free volume
fraction. Moreover, the $z_p$-dependence is hardly noticeable. We
thus conclude that equation (\ref{eqalfa}) is accurate for
short-ranged screened-Coulomb repulsions for $q=0.1$ or,
equivalently, that the effect of the screened-Coulomb repulsion on
$\alpha$ is negligible.

We can now analyse the effect of the screened-Coulomb repulsion on
the phase behaviour. The osmotic pressures $\Pi$ and colloid
chemical potentials $\mu_c$ can be found by differentiation of the
Helmholtz free energy (\ref{eqomegamodel1}). The colloid volume
fractions in each of the coexisting phases, $\eta_c^1$ and
$\eta_c^2$, are obtained by equating $\mu_c$ and $\Pi$ at fixed
polymer reservoir concentration $\phi_p^r$. The Helmholtz free
energy density (\ref{Hallfeq}) is used for the solid phase, while
equation (\ref{CarnStarfeq}) is employed for the fluid. The free
volume theory predicts for the AOV model, i.e., a system of
colloidal hard spheres and ideal polymer, a broadening of the
fluid-solid transition with increasing $\phi_p^r$ for size ratios
$q<0.3$, while the fluid-fluid transition is metastable with
respect to a broad fluid-solid transition
\cite{Lekkerkerker1992,Dijkstra1999}.

\begin{figure}[ht]
 \begin{center}
  \includegraphics{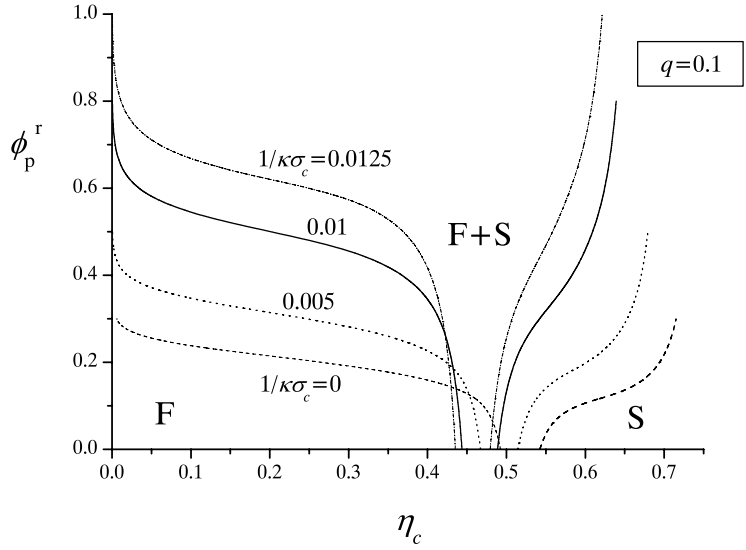}
   \end{center}
 \caption{Phase diagram of a mixture of charged  spheres and
 ideal polymer as obtained from free volume theory for
$q=0.1$ as a function of the colloid volume fractions $\eta_c$ and
the ideal polymer reservoir concentration $\phi_p^r$. The
screened-Coulomb repulsion $(\ref{Yukint})$ is characterised by
$\beta \epsilon = 20$ and various values of $\kappa \sigma_{c}$ as
indicated. $F$ and $S$ denote the stable fluid and solid fcc
phase. $F+S$ denotes the stable fluid-solid coexistence region. }
\label{fig5}
\end{figure}

We now turn to our case of charged colloidal spheres and ideal
polymer chains. In figure \ref{fig5}, we plot the phase diagram using the
free volume theory (\ref{eqomegamodel1}) for a size ratio $q=0.1$
and $(\kappa\sigma_c)^{-1}$ = 0.0,  0.005, 0.01, and  0.0125 in
the ($\eta_c, \phi_p^r$) plane. For $(\kappa\sigma_c)^{-1}=0$ and
$\phi^{r}_{p}=0$, we recover the well-known pure hard sphere
freezing transition at $\eta_{c}=0.494$ and $0.545$
\cite{Hoover1968}. In the case of charged spheres, the freezing
transition at $\phi^{r}_{p}=0$ shifts to lower colloid volume
fractions $\eta_{c}$, which is in line with the results in figure
\ref{fig3}, and is due to a larger effective volume of the charged spheres.
Figure \ref{fig5} shows clearly that the fluid-solid transition widens upon
increasing the polymer concentration. More specifically, the
broadening of the freezing transition shifts to higher $\phi_p^r$
with increasing range of the screened-Coulomb repulsion $(\kappa
\sigma_{c})^{-1}$. This can be explained as follows. Upon
increasing the range of the screened-Coulomb repulsion, $\eta_e=m
\eta_{c}$, and hence $f_c$, increases. At the same time the free
volume fraction $\alpha$ is not affected significantly (see figure
\ref{fig4}) upon adding a screened-Coulomb repulsion. So, in order to
attain a similar effect on $f$ (see Eq. (18)), a higher polymer concentration
is required to broaden the freezing transition.

\begin{figure}[ht]
 \begin{center}
  \includegraphics{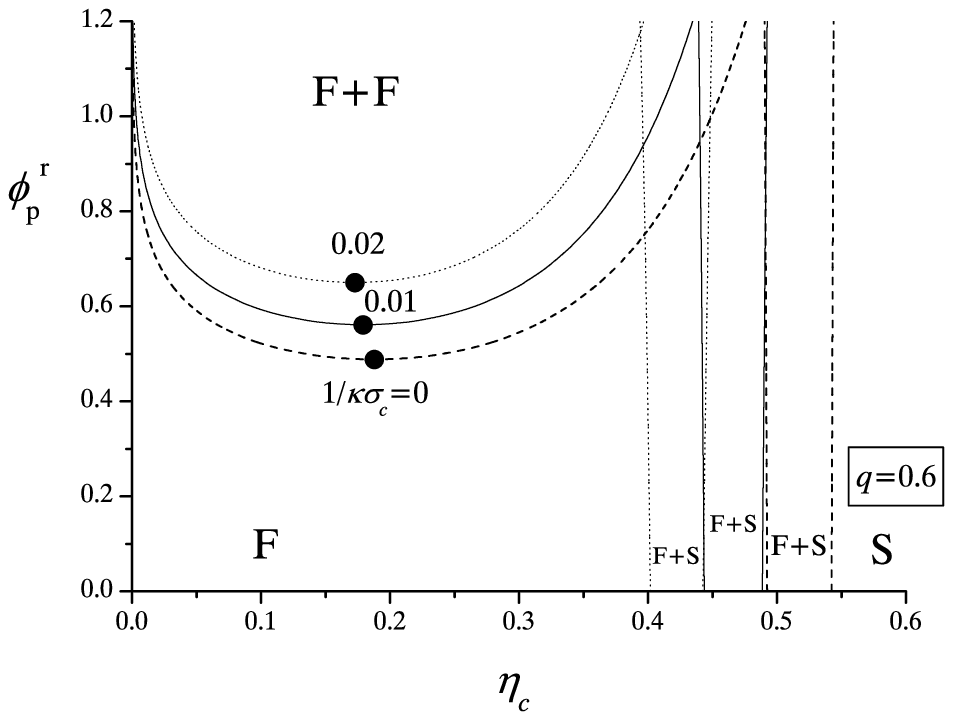}
   \end{center}
 \caption{Same as figure \ref{fig5} but for $q=0.6$. $F+F$ denotes the stable fluid-fluid coexistence
 region.}
\label{fig6}
\end{figure}

\begin{figure}[ht]
 \begin{center}
  \includegraphics{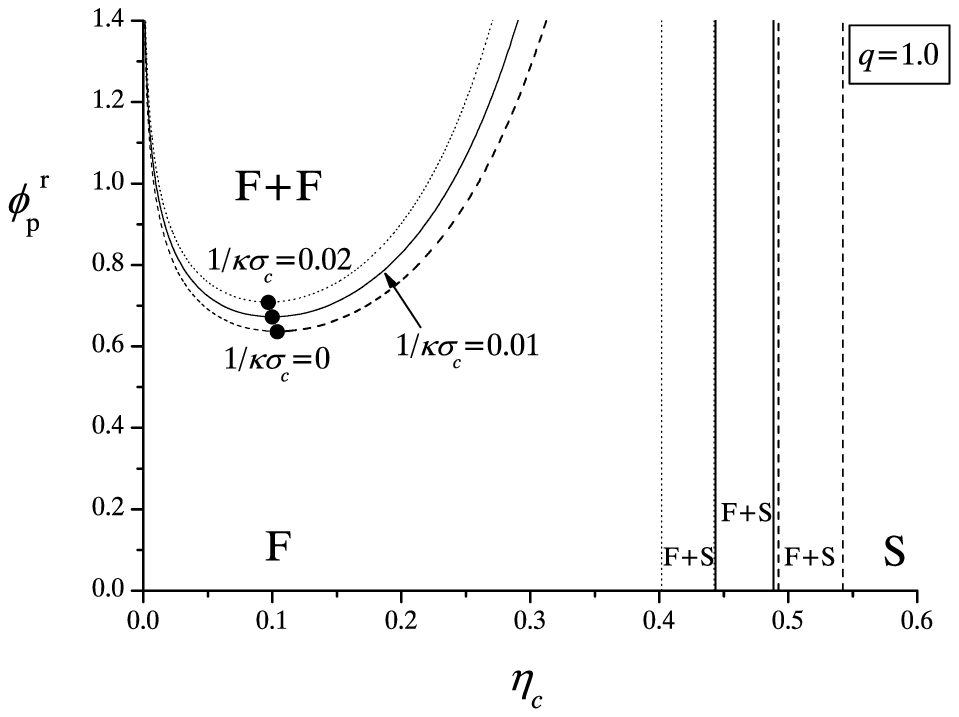}
   \end{center}
 \caption{Same as figure \ref{fig5} but for
$q=1.0$. $F+F$ denotes the stable fluid-fluid coexistence region.}
\label{fig7}
\end{figure}

For larger values of $q$, say $q>0.4$, a fluid-fluid coexistence
becomes stable in the AOV model, which dominates the phase
behaviour at colloid volume fractions $\eta_c<0.49$. In analogy
with figures \ref{fig1}b and \ref{fig1}c we choose $q=0.6$ and $1$ and study the
effect of the repulsive screened-Coulomb interaction on the
fluid-solid and fluid-fluid transition. In figures \ref{fig6} and \ref{fig7}, we
plot the predictions from free volume theory in the
($\eta_c,\phi_p^r$) representation for $q=0.6$ and $q=1.0$,
respectively.  We again find that  the freezing transition at
$\phi^{r}_{p}=0$ shifts to lower colloid volume fractions
$\eta_{c}$ upon increasing $(\kappa\sigma_c)^{-1}$. In addition,
figures \ref{fig6} and \ref{fig7} show  that the fluid-fluid demixing shifts to
higher $\phi_p^r$ with increasing range of the screened-Coulomb
repulsion $(\kappa \sigma_{c})^{-1}$. Hence, the trends are
similar as for $q=0.1$. The screened-Coulomb repulsion reduces the
depletion effect. The critical points are indicated as the filled
circles in figure \ref{fig6} and \ref{fig7}, and they indicate that the critical
colloid volume fraction $\eta_{c}$ shifts to somewhat smaller
values upon increasing the range of the soft repulsion.

\begin{figure}[ht]
 \begin{center}
  \includegraphics{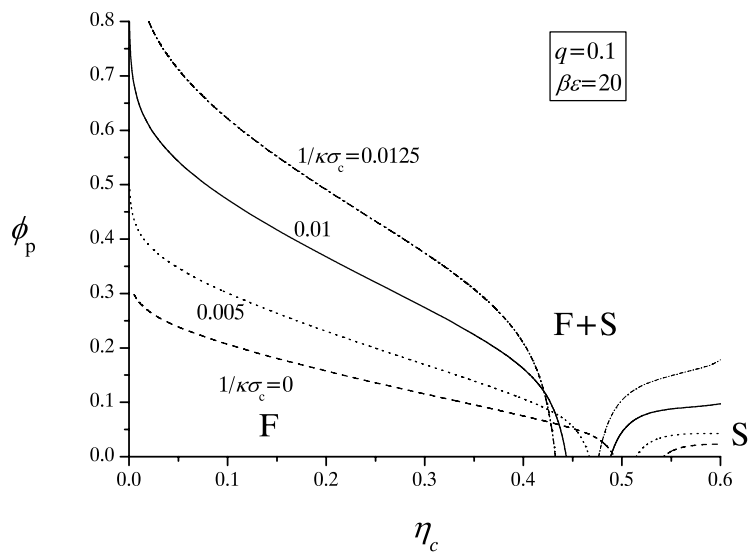}
   \end{center}
 \caption{Same as figure \ref{fig5} but as a function of the actual ideal polymer  concentration $\phi_p$.}
\label{fig8}
\end{figure}

\begin{figure}[ht]
 \begin{center}
  \includegraphics{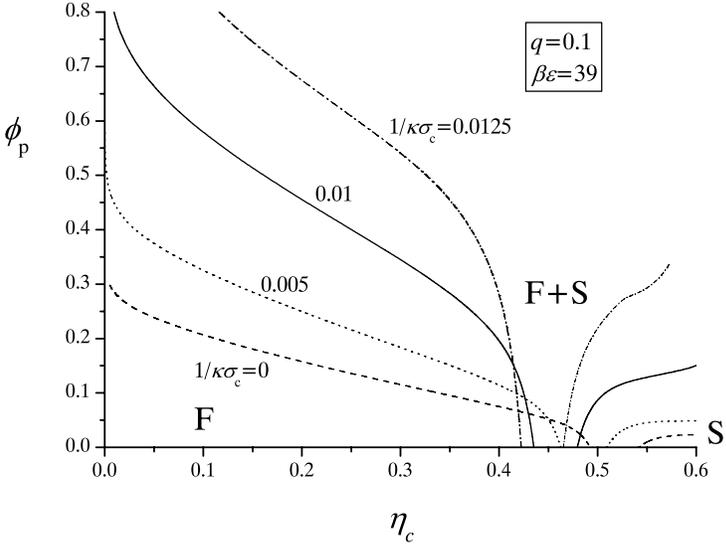}
  \end{center}
 \caption{Same as figure \ref{fig8}, but with $\beta \epsilon = 39$. }
\label{fig9}
\end{figure}

\begin{figure}[ht]
 \begin{center}
  \includegraphics{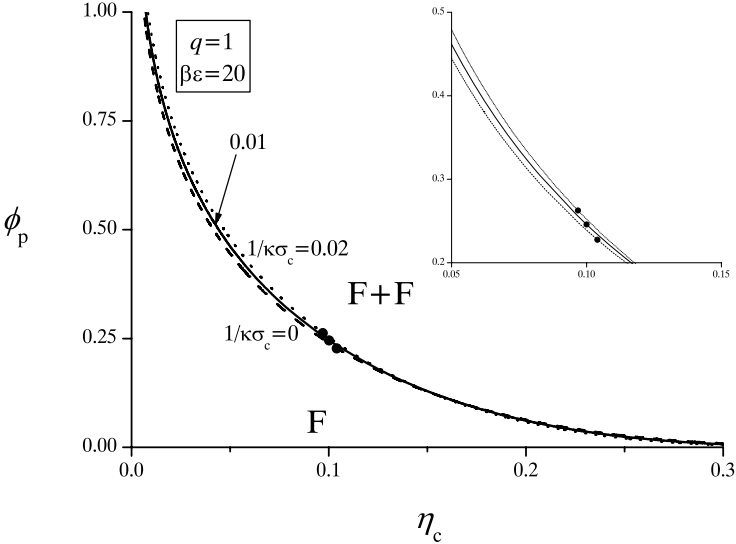}
   \end{center}
 \caption{Same as in  figure \ref{fig8}, but with size ratio $q=1.0$. 
 The filled circles indicate the location of the gas-liquid critical points. 
 The inset is a blow-up of the critical region of the liquid-gas binodal.}
\label{fig10}
\end{figure}

Finally, we convert the polymer reservoir concentration $\phi_p^r$
to that in the actual system $\phi_p$.  Figure \ref{fig8} shows the
conversion of the phase diagram of figure \ref{fig5} for $q=0.1$ and $\beta
\epsilon=20$ and Debye screening lengths $(\kappa
\sigma_{c})^{-1}=0$ (dashed curves; the pure hard sphere case),
$(\kappa \sigma_{c})^{-1}=0.005$ (dot-dashed curves), $(\kappa
\sigma_{c})^{-1}=0.01$ (full curves) and $0.0125$ (dotted curves).
The phase stability of a mixture of charged colloids and neutral
polymer chains in an aqueous salt solution is thus expected to
depend very sensitively on the screening length, and thus on the
salt concentration, at least for small size ratios $q$. In figure
\ref{fig9}, we investigate the effect of $\beta\epsilon$ on the phase
behaviour. We plot the converted phase diagram for the same set of
parameters as in figure \ref{fig8}, i.e., $q=0.1$ and varying Debye
screening lengths $(\kappa \sigma_{c})^{-1}$, but with
$\beta\epsilon=39$ instead of $\beta\epsilon=20$. There is a
striking similarity between the two sets of results and, hence, we
conclude that the effect of $\beta\epsilon$ is not significant.
For larger values of $\beta\epsilon$, the system becomes more
sensitive to $\kappa\sigma_c$ and thus  to the salt concentration.
Finally, figure \ref{fig10} shows the conversion of the phase diagram as
shown in figure \ref{fig7} for $q=1.0$ and $\beta \epsilon=20$ and Debye
screening lengths $(\kappa \sigma_{c})^{-1}=0$ (dashed curves; the
pure hard sphere case), $(\kappa \sigma_{c})^{-1}=0.01$ (full
curves) and $0.02$ (dotted curves).

\begin{figure}[ht]
 \begin{center}
  \includegraphics{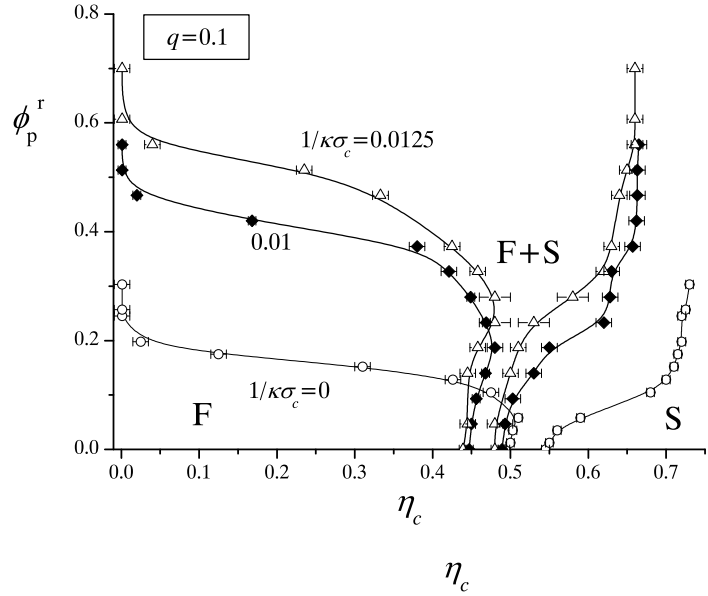}
   \end{center}
 \caption{Phase diagram of a mixture of charged  spheres and
 ideal polymer as obtained from simulations of
 the effective pair potential (\ref{totint}) for
$q=0.1$  as a function of the colloid volume fractions $\eta_c$
and the ideal polymer reservoir  concentration $\phi_p^r$. The
screened-Coulomb repulsion $(\ref{Yukint})$ is characterised by
$\beta \epsilon = 20$ and various values of $\kappa \sigma_{c}$ as
indicated.
 The curves serve as a guide to the eye. Open circles correspond to the pure
hard sphere case $(\kappa \sigma_{c})^{-1}=0$,
 closed circles to a screened-Coulomb repulsion with $(\kappa \sigma_{c})^{-1}=0.01$ and
open triangles to $(\kappa \sigma_{c})^{-1}=0.0125$. $F$ and $S$
denote the stable fluid and solid fcc phase. $F+S$ denotes the
stable fluid-solid coexistence region.}
\label{fig11}
\end{figure}

It follows that an increase of the reduced Debye length $(\kappa
\sigma_{c})^{-1}$ shifts the fluid-fluid coexistence curves
upwards. Using a PRISM approach, Ferreira \textit{et al}.
\cite{Ferreira2000} also found this trend  for the spinodal curve
of the demixing fluid (see their figure \ref{fig11}) based on determining
the composition where the inverse structure factor vanishes in the
long wavelength limit. Figures \ref{fig7} and \ref{fig10} show clearly that the
shift in polymer concentration of the fluid-fluid binodals for
$q=1$ is weak compared to the shift in the fluid-solid binodals
for small $q$.

A relevant quantity that measures the relative influence of the
screened-Coulomb repulsive pair interaction is $(\kappa
\sigma_{p})^{-1}$ or $(\kappa \sigma_{c} q)^{-1}$. Hence, the size
of the polymer chains (or the depletion thickness) as compared to
the range of the repulsion determines the relative importance of
the screened-Coulomb repulsion
 on the total \textit{effective} depletion interaction. In biological systems such as charged proteins mixed with neutral
polysaccharides, where often $\kappa^{-1} < \sigma_{c} q$, we expect that the
 phase behaviour is only moderately sensitive to the salt concentration.
Decreasing the salt concentration significantly is then expected
to stabilise the charged biocolloid dispersion against
depletion-induced demixing. This explains the enhanced miscibility
found in mixtures of proteins mixed with neutral non-adsorbing
polysaccharides in aqueous salt solutions \cite{Grinberg1997}. In
many charged colloidal dispersions the soft repulsion is expected
to suppress the depletion effect. In several applications such as
paints and food dispersions where colloidal particles are mixed
with polymer chains, a screened-Coulomb repulsion helps
stabilising the dispersion.

\subsection{Simulations using the effective
Hamiltonian}\label{simulations} In order to test the validity of
the predictions from free volume theory, we compare our results
with Monte Carlo simulations. We determine the phase diagram of
the effective one-component system by calculating the
dimensionless free energy density $f= \beta F v_c/V$ as a
function of the colloid packing fraction $\eta_c$ and the fugacity
of the polymer chains $z_{p}$ in simulations. For non-interacting
chains
 the fugacity $z_{p}$ equals the density of polymer chains $\rho_p^r$ in the
 corresponding reservoir.
As the free energy density cannot be measured directly in a Monte
Carlo simulation, we use thermodynamic integration to relate the
free energy of the effective system to that of a reference system
at the same colloid volume fraction $\eta_c$.
 To this end, we write the total free energy density as the sum of
two contributions
\begin{equation}
f(N_c,V,z_p) \;=\; f_{c}(N_c,V) +
f_{\mbox{\footnotesize{dep}}}(N_c,V,z_p) \ , \label{E:free-en}
\end{equation}
where  $f_{c}(N_c,V)$ is the free energy density  for a system of
$N_c$ charged colloids in a volume $V$ interacting with a
hard-core repulsive Yukawa potential  (\ref{Yukint}), and
$f_{\mbox{\footnotesize{dep}}}$ is the contribution of the
depletion potential (\ref{AO}) to the free energy density. The
free energy density $f_{c}(N_c,V)$ is computed using the so-called
$\lambda$-integration for the fluid phase  and the standard
Frenkel-Ladd method for the solid phase \cite{Frenkel2002} with
the Einstein crystal as a reference. A second
$\lambda$-integration is then carried out, for both the fluid and
the solid phase, to determine the free energy density contribution
$f_{\mbox{\footnotesize{dep}}}$. For more details of the
simulations, we refer the reader to Appendix 2. In order to map
out the phase diagram we determine the total free energy density
$f(\eta_{c},z_{p})$ for many state points $(\eta_{c},z_{p})$ in
simulations. We employ  common tangent constructions at fixed
$z_p$ to obtain the coexisting phases \cite{Dijkstra1999E}. In
order to assess the performance of the free volume theory, phase
diagrams were determined for some of the parameters as in  figure
\ref{fig5} ($q=0.1$). The results are shown in figure \ref{fig11} and agree
semi-quantitatively with those in figure \ref{fig5}. The main difference is
due to the fact that the results of both approaches deviate
already for the AOV model (hard sphere colloids with ideal
polymer, i.e., $(\kappa\sigma_c)^{-1}=0$)  \cite{Dijkstra1999}.
For the AOV model, the theoretical binodal is shifted with  a
factor of about 1.3 in $\phi_p^r$ compared to the simulations.
This factor between the theoretical predictions and the simulation
results is about 1.2 when the screened-Coulomb repulsion is added.

\begin{figure}[ht]
 \begin{center}
  \includegraphics{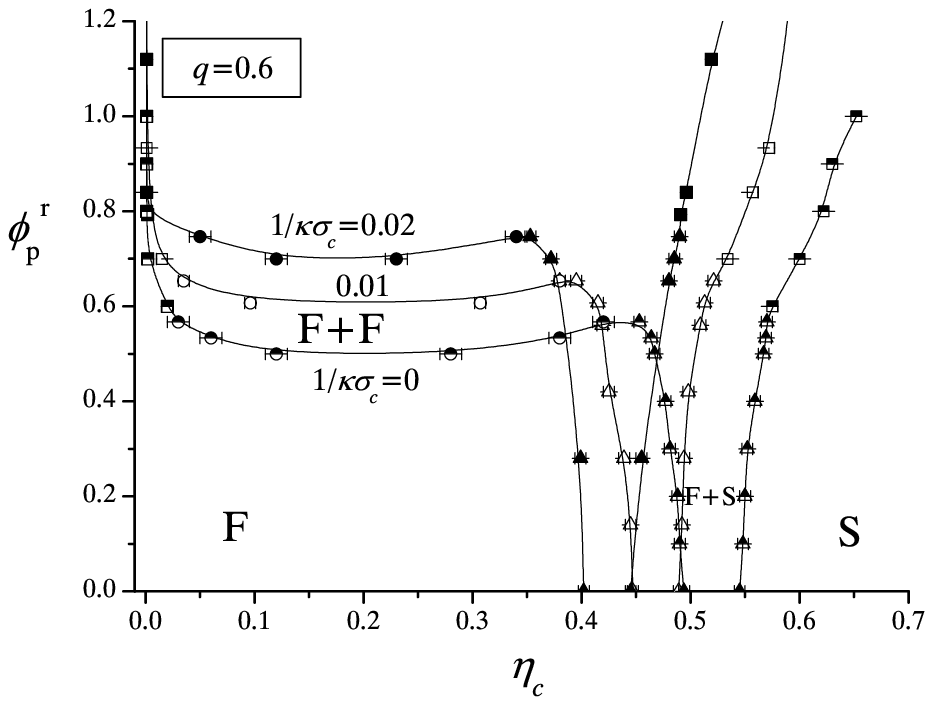}
 \label{fig12}
 \caption{Same as figure \ref{fig11}, but for
$q=0.6$. $F+F$ denotes the stable fluid-fluid coexistence region.}
 \end{center}
\end{figure}

\begin{figure}[ht]
 \begin{center}
  \includegraphics{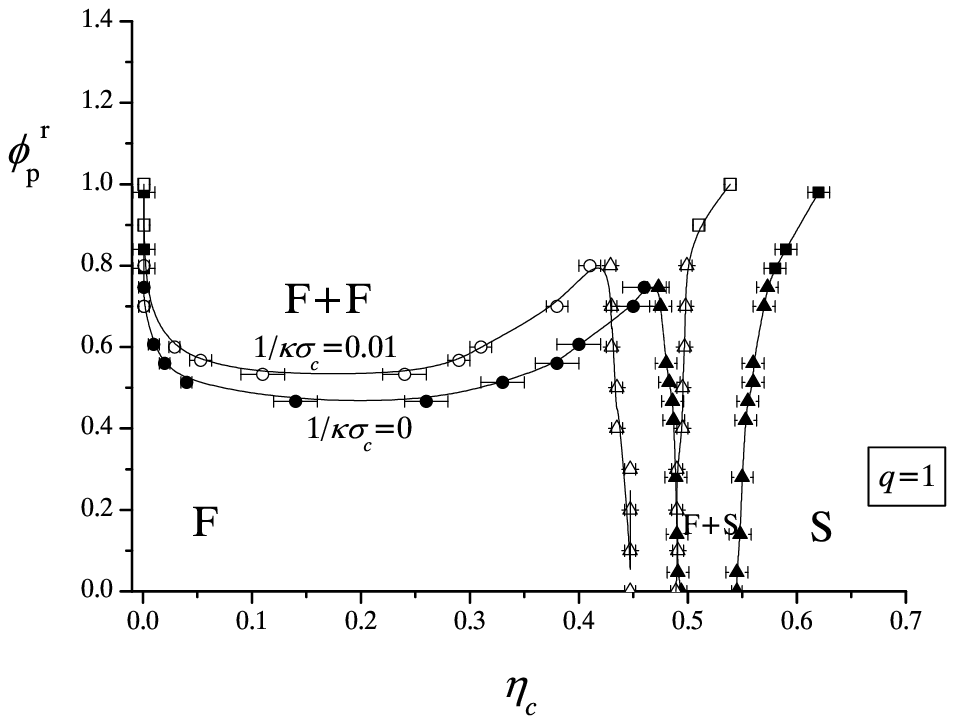}
 \label{fig13}
 \caption{Same as figure \ref{fig11}, but for
$q=1$. $F+F$ denotes the stable fluid-fluid coexistence region.}
 \end{center}
\end{figure}

We also compare the free volume theory results for the phase
behaviour  for $q=0.6$ and 1 as shown in figures \ref{fig6} and \ref{fig7} with
computer simulations. The phase diagrams obtained from Monte Carlo
simulations of the effective one-component systems are plotted in
figure \ref{fig12} ($q=0.6$) and figure \ref{fig13} ($q=1.0$). The main theoretical
trends are also found in our  simulation results. Again, the data
of the simulations suggest that a higher polymer concentration is
required to induce the fluid-fluid transition upon increasing the
range of the soft repulsion.

Finally, we stress that the free volume theory incorporates some
of the many-body effects which are present at large $q$, while our
simulations are based on a two-body approximation to the effective
Hamiltonian. It is therefore  difficult to make a direct
comparison between the simulation results and those obtained from
free volume theory. However, for $q\leq 0.1547$, the mapping of
the charged colloid-polymer mixture onto an effective
one-component Hamiltonian based on depletion pair potentials is
exact and thus a direct comparison is feasible for our results for
$q=0.1$.

\section{Concluding remarks}
We have studied the effect of a short-ranged screened-Coulomb
repulsion  on the  phase stability of mixtures containing charged
spheres and  non-adsorbing polymer chains. The charged spheres are
described as hard spheres with an additional screened-Coulomb or
Yukawa repulsion with the screening length given by the Debye
length $\kappa^{-1}$, setting the range of the soft repulsion. The
phase behaviour of the charged sphere dispersion is described
using standard expressions for the colloidal hard sphere fluid and
fcc crystal with the hard sphere volume fraction replaced by an
\textit{effective} volume fraction that depends on the Yukawa
interaction between the spheres.  Our results obtained from free
volume theory and Monte Carlo simulations show that the additional
screened-Coulomb repulsion reduces the depletion effect. For
mixtures of small polymers plus relatively large charged spheres
the fluid-solid transition is shifted to significantly larger
polymer concentrations with increasing Debye screening length
$\kappa^{-1}$, while for relatively larger polymers the effect is
weaker: the resulting fluid-fluid binodal is affected weakly by
adding a short-ranged soft repulsion. In general, the range of the
screened-Coulomb repulsion compared to the range of the depletion
attraction determines qualitatively  the reduction of the
depletion effect, and hence, shifts the fluid-fluid and
fluid-solid binodals correspondingly. In conclusion, a mixture of
charge-stabilised colloids and non-adsorbing polymers at large
concentrations of both components can be stabilised by lowering
the salt concentration, which increases the range of the
screened-Coulomb interaction of the colloids.
\\

\ack This work was supported by the SoftComp Network of
Excellence. We thank M. Schmidt, A.-P. Hynninen, H.N.W.
Lekkerkerker, A. Vrij, G. A. Vliegenthart and J. Buitenhuis for
useful discussions. We thank the Dutch National
Computer Facilities foundation for access to the SGI Origin 3800 and
SGI Altix 3700.
\\

\section*{Appendix 1: Free volume fraction in a charged sphere dispersion}
\label{appen1}

In this appendix we consider the free volume fraction $\alpha$
that is available for ideal polymer chains in a sea of charged
spheres with diameter $\sigma_c$. As the centre-of-mass of the
polymer chains is excluded from the centre-of-mass of the charged
colloids by a distance $(\sigma_c+\sigma_p)/2$  and the polymer
interactions are ideal, $\alpha$ is just the free volume fraction
for a single hard sphere with diameter $\sigma_p$ in a sea of
charged spheres. This free volume fraction $\alpha$ can be
determined from the chemical potential of the polymer chains. The
chemical potential for inserting a polymer in a sea of charged
spheres consists of an ideal gas term and a work term $W$.
\begin{equation}
\beta\mu_p = \ln \rho_p\Lambda_p^3 + W
\end{equation}
Following Widom's particle insertion method \cite{Widom1967}, the
required work to insert a polymer $W$ is equal to $\beta W=-\ln
\alpha$. The work $W$ can be determined from scaled particle
theory, that considers two limits. In this theory, the size of the
particle is scaled with a parameter $x$. In the limit $x\ll 1$,
the polymer coils reduce to points, and hence the volume fraction
available to the polymer is simply unity minus the sum of the
overall hard sphere volumes plus the depletion layers around them:
\begin{equation}
\label{eqalfasmall} \alpha(x\ll 1)\;=\;1 -
\frac{\pi}{6}\rho_{c}\left ( \sigma_{c}\;+\; x
\sigma_{p}\right)^{3}
\end{equation}
where $\rho_{c}=N_c/V$ is the number of colloidal spheres per
volume (related to $\eta_{c}$ via
$\eta_{c}=\pi\rho_{c}\sigma_{c}^{3}/6$). Hence, it follows
\begin{equation}
\label{eqwsmall} \beta W (x\ll 1)\;=\;-\ln \left [1 -
\frac{\pi}{6}\rho_{c}\left ( \sigma_{c}\;+\; x
\sigma_{p}\right)^{3} \right ]
\end{equation}
On the other hand, if $x\gg1$, the work required to insert a
polymer coil in a sea of charged spheres, is approximately the
work to create a hole with the size of the polymer coil, which is
equal to the volume of the polymer coil times the osmotic pressure
$\Pi_c$ of the dispersion of charged colloids:
\begin{equation}
W(x \gg 1) = \frac{\pi}{6} x^3\sigma_{p}^{3} \Pi_c. \end{equation}
In scaled particle theory,  $W(x\ll 1)$ is expanded  about $x=0$
up to order $x^2$ and $W(x\gg1)$ is added as the $x^3$ term.
\begin{equation}
\label{eqwseries} W( x )\;=\;W(x=0)\;+\;x \left ( \frac {\partial
W}{\partial x}\right)_{x=0}\; +\;\frac {1}{2}x^{2} \left ( \frac
{\partial^{2} W}{\partial x^{2}}\right)_{x=0}\;+\; \frac{\pi}{6}
x^3\sigma_{p}^{3} \Pi_{c}
\end{equation}
 Scaling the polymer coils to the
desired size by $x=1$, yields
\begin{eqnarray}
\label{eqwalgemeen} \beta W (x=1) \;&=&\;-\ln \left [1 - \eta_{c}
\right]\;+\;3q\gamma  \;+\; \frac{1}{2}\left(6q^{2}\gamma\;+\; 9q^{2}\gamma^{2} \right)\nonumber \\
& & \;+\; \frac{\pi}{6} \sigma_{p}^{3} \beta \Pi_{c}
\end{eqnarray}
where $\gamma=\eta_c/(1-\eta_c)$. Hence $\alpha$ follows
straightforwardly from $W\equiv W(x=1)$. For pure hard spheres one
usually takes
 the Percus-Yevick result for the pressure $\Pi_{HS}$  from the virial route
(see \cite{HansenMcDonald1986}) since it is consistent with SPT
\begin{equation}
\label{eqPYvirialP}
\frac{\beta \Pi_{HS}}{\rho_{c}}\;=\;\frac {1\;+\;\eta_{c}\;+\;\eta_{c}^{2}}{\left ( 1\;-\;\eta_{c}\right)^{3}}\;=\;
\frac {1}{1\;-\;\eta_{c}}\;
+\;\frac {3\eta_{c}}{\left ( 1\;-\;\eta_{c}\right)^{2}}\;+\;
\frac {3\eta_{c}^{2}}{\left ( 1\;-\;\eta_{c}\right)^{3}}
\end{equation}
Inserting this expression for $\Pi_{HS}$ into equation
\ref{eqwalgemeen}
 for $\Pi_{c}$ yields
\begin{equation}
\label{eqwHS} \beta w\;=\;-\ln \left [1 - \eta_{c}
\right]\;+\;(A\;+\;C)\gamma\;+\; (B\;+\;3C)\gamma^{2}\;+\;
3C\gamma^{3}
\end{equation}
where $A$, $B$, and $C$,  are defined below equation \ref{eqalfa}.
Hence, we arrive at the standard SPT result for the free volume
fraction of ideal polymer in a sea of hard spheres
\cite{Lekkerkerker1992}
\begin{equation}
\label{eqalfaHS} \alpha\;=\;\left (1 - \eta_{c} \right) \exp
\left\lbrack-\left((A\;+\;C)\gamma\;+\; (B\;+\;3C)\gamma^{2}\;+\;
3C\gamma^{3} \right) \right\rbrack
\end{equation}

In case of colloidal spheres interacting with a Yukawa pair
potential, we rewrite equation (\ref{eqPYvirialP})  following the
approach outlined in section 3 giving a  pressure
\begin{equation}
\label{eqPYvirialPsoft} \frac {\beta
\Pi_{Yuk}}{\rho_{c}}\;=\;\frac
{1\;+\;\eta_{e}\;+\;\eta_{e}^{2}}{\left (
1\;-\;\eta_{e}\right)^{3}}\;=\; \frac {1}{1\;-\;\eta_{e}}\;
+\;\frac {3\eta_{e}}{\left ( 1\;-\;\eta_{e} \right)^{2}}\;+\;
\frac {3\eta_{e}^{2}}{\left ( 1\;-\;\eta_{e} \right)^{3}}
\end{equation}
where $\eta_e$ is defined by equation (\ref{eff vol frac}). Using
this expression for $\Pi_{c}$ in equation
\ref{eqwalgemeen} yields equation \ref{eqalfa}.\\

\section*{Appendix 2: Technical details of the simulations} \label{appen2}
This section describes the technical details of the simulations.
To this end we consider the total effective one-component
Hamiltonian of the colloids at fixed polymer fugacity $z_p$
\begin{equation}
H(z_p)= H_{cc}\;+\;H_{\mbox{\footnotesize{dep}}}(z_p) \,
\end{equation}
where the  Hamiltonian $H_{cc}$ consists of a sum of
colloid-colloid pair potentials $U_{cc}$, which can be split into
a sum of hard-sphere potentials $U_{HS}$ and a sum of repulsive
Yukawa potentials:
\begin{eqnarray}
H_{cc}=\sum_{i<j}^{N_c} U_{cc}(R_{ij})=\sum_{i<j}^{N_c}
U_{HS}(R_{ij}) + \sum_{i<j}^{N_c} U_{Yuk}(R_{ij})
\end{eqnarray}
with \begin{equation} U_{Yuk}(R_{ij}) = \epsilon \frac{\exp(\kappa
\sigma_c(R_{ij}/\sigma_c-1))}{R_{ij}/\sigma_c}
\hspace{1cm}\mbox{for} \hspace{1cm}R_{ij}>\sigma_c \end{equation}
 and
$H_{\mbox{\footnotesize dep}}$ a sum of depletion potentials
$U_{\mbox{\footnotesize dep}}$ (\ref{AO}):
\begin{equation}
H_{\mbox{\footnotesize dep}}(z_p)\;=\; \sum_{i<j}^{N_c}
U_{\mbox{\footnotesize dep}}(R_{ij};z_p).
\end{equation}
The different Hamiltonians give rise to the corresponding free
energy contributions of equation (\ref{E:free-en}).

To determine  the free energy contribution of the Yukawa
potential, for the fluid phase, we introduce the auxiliary
Hamiltonian
\begin{equation}
H^{fluid}_{cc,\lambda} \;=\; \sum_{i<j}^{N_c} U_{HS}(R_{ij}) +
\lambda \sum_{i<j}^{N_c} U_{Yuk}(R_{ij})
\end{equation}
where $0 \leq \lambda \leq 1$ is a dimensionless coupling
parameter: at $\lambda=0$ the auxiliary Hamiltonian is that of a
pure system of $N_c$ hard spheres, while at $\lambda=1$ it is the
Hamiltonian of $N_c$ charged spheres. The free energy is
determined by applying the standard $\lambda$-integration
technique \cite{Frenkel2002}
\begin{equation}
 f_{c}(N_c,V)\;=\; f_{HS}(N_c,V,\lambda=0)\;+\;\frac{v_c}{V}
 \int_0^1 d \lambda \left \langle  \sum_{i<j}^{N_c} \beta U_{Yuk}(R_{ij}) \right\rangle_{N_c,V,\lambda} .
\end{equation}
The angular brackets denote a canonical average over a system of
$N_c$ particles interacting with the Hamiltonian
$H^{fluid}_{cc,\lambda}$, while $f_{HS}(N_c,V,\lambda=0)$ is the
free energy of a system of hard spheres, for which we use the
Carnahan-Starling expression\ \cite{Carnahan1969}. We start the
canonical simulations from a random, non-overlapping
configuration, and use 15000 MC sweeps per particle for
equilibration and typically 15000  production moves for each value
of the coupling parameter $\lambda$. In principle, the free energy
of the solid phase can be computed with the same technique using
the Hall expression  for the free energy of the hard sphere
crystal \cite{CKHall1972}. However the latter is only properly
defined for packing fractions  larger than the  value at hard
sphere freezing  $\eta_{c}=0.545$, while charged spheres can yield
crystal phases at  lower packing fractions. A different technique
is used for the solid phase by introducing the auxiliary
Hamiltonian
\begin{equation}
H^{solid}_{cc,\lambda} \;=\; \sum_{i<j}^{N_c} U_{cc}(R_{ij}) \;+\;
\lambda k_BT \sum_{i=1}^{N_c} ({\bf r}_i-{\bf
r}_{o,i})^2/\sigma_c^2 \ ,
\end{equation}
where ${\bf r}_{o,i}$ denote the ideal lattice position of
particle $i$ in a  fcc crystal. The free energy is computed by
applying the integration technique introduced by Frenkel and
co-workers \cite{Frenkel1984,Polson2000}
\begin{eqnarray}
 f_{c}(N_c,V)\;&=&\; f_{ein}^{CM}(N_c,V,\lambda=\lambda_{max})+f_{corr}(N_c,V)\nonumber \\
 & & \;-\;\frac{v_c}{V}
 \int_0^{\lambda_{max}} d \lambda \left \langle   \sum_{i=1}^{N_c} ({\bf r}_i-{\bf r}_{o,i})^2/\sigma_c^2 \right \rangle_{\lambda}^{CM} ,
\end{eqnarray}
where the angular brackets denote a canonical average of the mean
square displacement of $N_c$ particles interacting with the
Hamiltonian $H^{solid}_{cc,\lambda}$, while the superscript $CM$
 denotes that it is calculated for a crystal with fixed
centre of mass. The parameter $\lambda_{max}$ is chosen such that
for $\lambda=\lambda_{max}$ the system behaves like a
non-interacting Einstein crystal with  fixed centre of mass and
Madelung energy $U_{Yuk}({\bf r}_{0}^{N_c})$, i.e., the potential
energy of a crystal with all particles at their ideal lattice
positions. Typical values for $\lambda_{max}$ range from $1000$ to
$100000$ for high densities. The free energy of a non-interacting
Einstein crystal with fixed centre of mass reads
\begin{eqnarray}
f_{ein}^{CM}(N_c,V,\lambda=\lambda_{max})&=&\frac{v_c}{V} \beta
U_{Yuk}({\bf r}_{0}^{N_c}) - \frac{3(N_c-1)v_c}{2V} \ln \left
\lbrack\frac{\pi}{\lambda_{max}}\right \rbrack \nonumber \\& &+
\frac{(N_c-1)v_c}{V}\ln\left \lbrack\frac{\Lambda_c^3}{\sigma_c^3}
\right \rbrack
\end{eqnarray}
The correction term $f_{corr}$ arises when the constraint on the
centre of masses is released, i.e., the Helmholtz free energy
difference between the unconstrained and constrained crystal:
\begin{equation} f_{corr}(N_c,V)= \frac{v_c}{V}\ln \left \lbrack
\frac{\Lambda^3}{VN_c^{1/2}}\right \rbrack.\end{equation}  The
equilibration is done for 15000 MC steps per particle, and the
averages are taken for 15000 MC steps per particle.

To determine the free energy contribution of the AOV depletion
potential $f_{\mbox{\footnotesize dep}}$ we employ a second
thermodynamic integration for the solid and fluid phase:
\begin{eqnarray}
f_{\mbox{\footnotesize dep}}&=&f(N_c,V,\rho_p^r) -
f(N_c,V,\rho_{p}^r=0)\nonumber \\
&=&\int_0^{\rho_{p}^r} d \rho_p^r{'} \left ( \frac{\partial
f(N_c,V,\rho_p^r{'})}{\partial \rho_p^r{'}}\right
)_{N_c,V,\rho_{p}^r{'}}\label{E:integral}
\end{eqnarray}
The system at $\rho_p^r=0$ is a  system of  colloids interacting
with pair potentials $U_{cc}$. The integral is calculated by
dividing the interval $[0,\rho_{p}^r]$ in 20 to 30 equally spaced
intervals. We used 10000 MC steps per particles for equilibration,
while averages were taken for  20000 MC steps per particle. In
addition we used the integrand  to determine the number density of
ideal polymer
 in the system using the thermodynamic relation
\begin{equation}\label{alpha}
 \left (\frac{\partial f(N_{c},V,\rho_{p}^r)}{\partial \rho_{p}^r}\right ) =
 \left (\frac{\partial f(N_{c},V,z_{p})}{\partial z_{p}}\right )
 =-v_c\frac{ \langle \rho_{p} \rangle_{\rho_{p}^r}}{\rho_p^r}
\end{equation}

\section*{References}
\bibliographystyle{unsrt}
\bibliography{RTrefs}

%%%%%%%%%%
\clearpage


\end{document}